# Integer Quantum Hall Effect in Graphene Channel with p-n Junction at Ferroelectric Substrate Domain Wall

*Maksym V. Strikha[1,2*], Anatolii I. Kurchak[1], and Anna N. Morozovska[3†],*

[1] *V.Lashkariov Institute of Semiconductor Physics, National Academy of Sciences of Ukraine,*
*pr. Nauky 41, 03028 Kyiv, Ukraine*

[2] *Taras Shevchenko Kyiv National University, Radiophysical Faculty*
*pr. Akademika Hlushkova 4g, 03022 Kyiv, Ukraine*

[3] *Institute of Physics, National Academy of Sciences of Ukraine,*
*pr. Nauky 46, 03028 Kyiv, Ukraine*

**Abstract**

We revealed that 180-degree ferroelectrics domain walls (FDWs) in a ferroelectric substrate, which induce p-n junctions in a graphene channel, lead to the nontrivial temperature and gate voltage dependences of the perpendicular and parallel modes of the integer quantum Hall effect. In particular the number of perpendicular modes $v_\perp$, corresponding to the p-n junction across the graphene channel varies with gate voltage increase from small integers to higher fractional numbers, e.g. $v_\perp$=1, 1.5, 1.6(6), 1.75, 1.9, 2,…, 5.1, 6.875, …9.1…, 23,… in the vicinity of the transition from ferroelectric to paraelectric phase. These numbers and their irregular sequence principally differ from the sequence of fractional numbers $v = 3/2, 5/3…$ reported earlier. The origin of the unusual $v_\perp$-numbers is significantly different numbers of the edge modes, $v_1$ and $v_2$, corresponding to significantly different concentration of carriers in the left ($n_1$) and right ($n_1$) domains of p-n junction boundary. The concentrations $n_1$ and $n_2$ are determined by the gate voltage and polarization contributions, and so their difference originates from the different direction of spontaneous polarization in different domains of ferroelectric substrate. The difference between $n_1$ and $n_2$ disappears with the vanishing of spontaneous polarization in a paraelectric phase. The phase transition from the ferroelectric to paraelectric phase can take place either with the temperature increase (temperature-induced phase transition) or with the decrease of ferroelectric substrate thickness (thickness-induced phase transition).

---
[*] e-mail: maksym.strikha@gmail.com
[†] e-mail: anna.n.morozovska@gmail.com

The unique graphene band spectrum [1, 2, 3] leads to unconventional integer quantum Hall effect (**IQHE**) [4, 5, 6]. It had been demonstrated theoretically, that Dirac-like spectrum of graphene, and, consequently, additional double degeneration of a zero Landau level (**LL**), which is common for conduction and valence bands, results in a special form of a Hall quantization [7]:

$$\sigma_{xy} = -\frac{e^2}{2\pi\hbar}\nu, \qquad \nu = \pm 2(2k+1). \qquad (1)$$

Where $\sigma_{xy}$ is the xy-component of conductance tensor. The value $\nu$ is called the number of edge modes, see [6]. The Hall plateaus are centered around the values of $\nu = \pm 2(2k+1)$, where $k = 0, 1, 2...$ is an integer. The internal spin-valley symmetry in graphene leads to four-fold degeneracy of each LL, but LL with $k = 0$ has an additional double degeneracy. Nonzero k numbers are given by expression [6]:

$$k = \left[\frac{n}{4n_B} - \frac{1}{2}\right]. \qquad (2)$$

Symbol "[]" stands for the integer part of a number. 2D concentration of electrons in graphene channel is $n$ and $n_B = eB/(2\pi\hbar)$ is the density of magnetic field flux, threading the 2D surface corresponding to the degree of the $k$-th LL occupation in traditional terms of IQHE.

In Ref.[8] authors have explained theoretically the peculiarities of IQHE, observed experimentally in graphene with p-n junction across the conduction channel [9]. It has been demonstrated, that in the bipolar regime the electron and hole modes can mix at the p-n boundary, leading to current partition and quantized short noise. On the contrary, very recently in Ref.[10] the formation of IQHE with p-n junctions created along the longitudinal direction of graphene cannel had been studied, and the enhanced conductance can be observed in the case of bipolar doping. In both cases (p-n junction along and across the channel) IQHE observation can be exploited to probe the behavior and interaction of quantum Hall channels.

In [11, 12] we have studied the conductivity of graphene channel with p-n junction, induced by a 180-degree ferroelectric domain wall (**FDW**) in ferroelectric substrate. Experimentally p-n junction in graphene at FDW was studied in Refs. [13, 14]. These papers gave an evidence for a possibility to get p-n junction in graphene without complicated system of several local gates, local doping etc. In [15] we have studied p-n junction dynamics induced in a graphene channel by ferroelectric-domain motion in the substrate. We have demonstrated a possibility how to vary a number of p-n junctions in a channel between the source and drain electrodes by the motion of FDW in ferroelectric substrate.

In this paper we study the peculiarities of IQHE in graphene channel with p-n junction at FDW in a substrate, taking into account the FDWs for reasons, examined in Refs.[11, 12, 15]. System under consideration is similar to one examined in Ref.[15] and shown in **Fig.1(a)**. It includes top gate electrode, oxide dielectric layer, graphene conducting channel with source and drain electrodes, ferroelectric substrate and bottom electrode. The strong magnetic field **B** is applied normally to the graphene channel plane. The spontaneous polarization changes its direction from $-P_S$ to $+P_S$ on different sides of the 180-degree FDW. The FDW induces p-n junction in graphene at that the wall plane that coincides with the p-n junction position (see Refs. [11, 12, 15] for details).

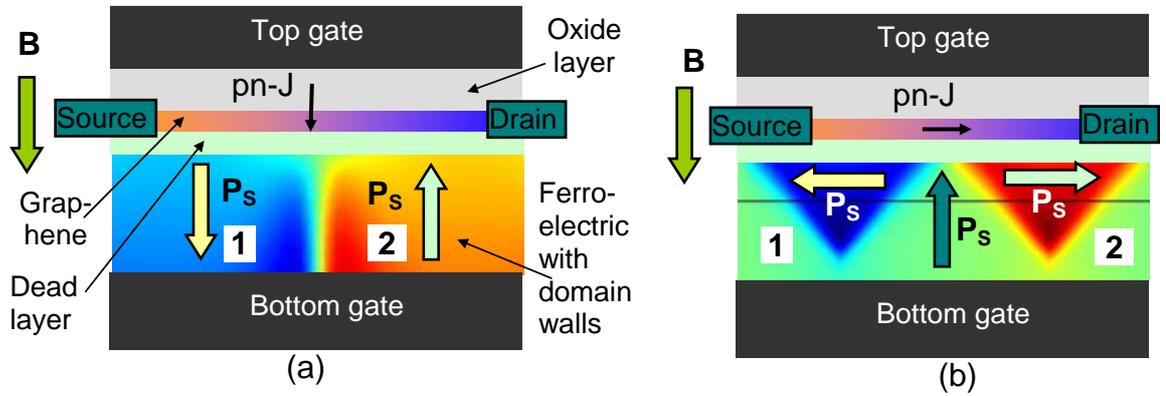

**Figure 1.** (a) Schematics of the considered heterostructure "top gate/ dielectric oxide layer / graphene channel / ultra-thin paraelectric dead layer/ ferroelectric substrate with a 180-degree domain wall / bottom gate". The spontaneous polarization changes its direction from $-P_S$ to $+P_S$ on different sides of the FDW. The strong magnetic field **B** is applied normally to the graphene channel plane**. (b)**. Schematics of the p-n junction along the graphene conducting channel created by in-plane flux-closure domains.

The concentration of electrons in a graphene channel placed at a single Ising-type 180-degree FDW is governed by different expressions [11], corresponding to the left (subscript "1") and to the right (subscript "2") sides of the wall,

$$n_1 = \frac{\varepsilon\varepsilon_0 V_g}{ed} + \frac{P_S}{e}, \qquad n_2 = \frac{\varepsilon\varepsilon_0 V_g}{ed} - \frac{P_S}{e}. \qquad (3)$$

Here $V_g$ is top gate voltage, $\varepsilon_0$ is dielectric permittivity of vacuum, $\varepsilon$ is relative dielectric permittivity of oxide dielectric layer, and $d$ is the layer thickness. The values $n_1$ and $n_2$ are taken far from the immediate vicinity of the domain wall, where polarization profile gradually changes its value from $-P_S$ to $+P_S$. This is possible because the thickness of 180-degree FDW (~several nm) is much smaller than the graphene channel length.

Corresponding number of edge modes, $\nu_{1,2}$, with carriers concentrations to the left and to the right sides of p-n junction boundary $n_{1,2}$, are

$$\nu_1 = \pm 2(2k_1 + 1), \qquad k_1 = \left[\frac{n_1}{4n_B} - \frac{1}{2}\right], \qquad (4a)$$

$$\nu_2 = \pm 2(2k_2 + 1), \qquad k_2 = \left[\frac{n_2}{4n_B} - \frac{1}{2}\right]. \qquad (4b)$$

Expressions (3)-(4) are valid until the spontaneous polarization exists, i.e. below the temperature $T_C$ of the ferroelectric-paraelectric phase transition, because FDW and induced p-n junction disappear at temperatures $T > T_C$.

Later we shall examine two cases: **(a)** p-n junction across the graphene conducting channel created by 180-degree domains [**Fig.1(a)**]; **(b)** p-n junction along the graphene conducting channel, [**Fig.1(b)**], created by in-plane flux-closure domains [16, 17]. For the **"perpendicular" case (a)** conductance in QHE regime is governed by a number of modes [8]

$$\nu_\perp = \frac{|\nu_1||\nu_2|}{|\nu_1| + |\nu_2|}, \qquad (5a)$$

where $\nu_{1,2}$ are determined by Eqs.(4) with different polarization directions ($\pm P_S$) corresponding to left and right sides of p-n junction.

On the contrary, for the **"parallel" case (b)** conductance in QHE regime is governed by a number of modes:

$$\nu_{\uparrow\uparrow} = |\nu_1| + |\nu_2|. \qquad (5b)$$

Physically Eqs (5a) and (5b) correspond to serial and parallel connection of domains between the source and drain electrodes. For the case of the high spontaneous polarization, when the concentration $P_S/e$ is much greater than the concentration induced by the top gate electrode $\varepsilon\varepsilon_0 V_g/ed$ for all the realistic values of parameters below the oxide layer breakdown, Eqs.(4) obviously yield $|\nu_1| \cong |\nu_2|$. This means that $\nu_\perp \cong \frac{1}{2}|\nu_1| \cong \frac{1}{2}|\nu_2|$, $\nu_{\uparrow\uparrow} \cong |2\nu_1| \cong 2|\nu_2|$ according to Eqs.(5). Therefore the conductance of graphene channel in IQHE regime in the case of the presence of p-n junction induced by FDW is suppressed in two times for serial connection geometry, and enhanced in two times for the parallel connection geometry. Therefore while IQHE without p-n junction is characterized by $\nu = 2, 6, 10...$, we get $\nu_\perp = 1, 3, 5...$ for the first case, and $\nu_{\uparrow\uparrow} = 4, 12, 20...$ for the second one. Note, that the fractional values of $\nu = 3/2, 5/3...$, predicted in Ref. [8] for serial connection and observed experimentally in Ref. [9], cannot be realized in the case $|P_S/e| \gg |\varepsilon\varepsilon_0 V_g/ed|$.

However for the case of comparatively small $P_S$, when the two addends in Eqs.(3) become of the same order of values, fractional values $\nu = 3/2, 5/3…$, predicted in Ref. [8] for serial connection, can be realized along with many others fractions (although only integer values for $\nu_{\uparrow\uparrow}$ are possible for the parallel connection).

For the case, when the inequality

$$|V_g| \geq d\frac{|P_S|}{\varepsilon\varepsilon_0}, \qquad (6)$$

can be reached without the breakdown of the oxide layer, we get the boundary between unipolar regions with different level of doping instead of p-n junction. As it was demonstrated in Ref. [8], the conductance across the boundary is determined by those edge modes with backscattering suppressed in IQHE, which are in contact with both reservoirs. These yields to

$$\nu_\perp \cong \min\{|\nu_1|, |\nu_2|\}, \qquad (7a)$$

On the contrary, for the parallel connection the conduction is determined by a number of biased channels [10], and

$$\nu_{\uparrow\uparrow} \cong \max\{|\nu_1|, |\nu_2|\}. \qquad (7b)$$

The fulfillment of the condition (6) for a gate voltage $V_g = V_g^{cr}$ (where $V_g^{cr} = \pm d\frac{|P_S|}{\varepsilon\varepsilon_0}$) should lead to the minimums with plateaus on the chart of the corresponding modes (as it will be illustrated below).

The reasonable ranges of parameters in Eqs.(4)-(5) are magnetic field strength $B = (0.5-10)$ Tesla, gate voltage $|V_g| = (0-10)$ V, oxide layer thickness $d = (5-50)$ nm and its relative permittivity $\varepsilon = (4-300)$. Universal dielectric constant $\varepsilon_0 = 8.85\times10^{-12}$ F/m, electron charge $e = 1.6\times10^{-19}$ C and reduced Plank constant $\hbar = 1,054\times10^{-34}$ J s. At that small values $\varepsilon = (1-10)$ correspond to dielectrics (e.g. $SiO_2$) or dead layers formation at the ferroelectric surface, $\varepsilon = 10 - 80$ corresponds to high-k dielectrics ($HfO_2$, $TiO_2$) and $\varepsilon = (100-300)$ is possible for quantum paraelectrics like $SrTiO_3$. The temperature dependence of polarization, $P_S = P_0\sqrt{1-\frac{T}{T_C}}$ corresponds to the second order phase transition at Curie temperature $T = T_C$. The value $P_0$ can very in the range $(0.1-1)$ C/m$^2$, where small values corresponds to

ferroelectric relaxors (e.g. PVDF) and layered perovskites (e.g. $CuInP_2S_6$ and $CuInP_2Se_6$), and high values are characteristic for proper ferroelectrics (e.g. $LiNbO_3$, $PbTiO_3$, $BiFeO_3$). Notably that Curie temperature $T_C$ can vary from low temperatures (around (150-293)K for $CuInP_2(S,Se)_6$) to elevated temperatures (up to 1100 K for $BiFeO_3$) depending on the ferroelectric material. Importantly that $T_C$ can be made close to the room temperature (around 300 K) for proper ferroelectrics by decreasing the film thickness below 100 nm due to the existence of the thickness-induced phase transition into a paraelectric phase [18, 19].

Using above parameters one can estimate that the number of perpendicular modes $\nu_\perp$ can vary from 1 in the vicinity of $T_C$, where the condition (6) is valid to very high values (>$10^2$) far from the temperature. When the drift term $\frac{V_g}{d}$ approaches the polarization term $\pm\frac{|P_S|}{\varepsilon\varepsilon_0}$ in Eq.(6) the approximate equalities in Eqs.(7) become valid.

Actually two sharp symmetric stepped minimums are clearly seen at the voltage dependences of the modes $|\nu_1|$, $|\nu_2|$ and $\nu_\perp$ in **Fig.2(a)-(c)**. The voltage position and the distance between the minimums significantly decrease from 4V to 1V with the difference $\Delta T = T_C - T$ decrease from 20 K to 1 K [compare **Fig.2(a), (b)** and **(c)**]. In the vicinity of each minima either $\nu_\perp \cong |\nu_1|$ or $\nu_\perp \cong |\nu_2|$ (red curve almost coincide either with black one or with blue one). However the numbers of $|\nu_1|$ or $|\nu_2|$ are integers, 2, 4, 6, etc, the number of perpendicular modes $\nu_\perp(V_g)$ appears fractional and its denominator differs from 2 or 3 with increase of either $|\nu_1|$ or $|\nu_2|$ corresponding to the increase of the voltage difference $|V - V_g^{cr}|$ (see the difference between red, black or blue curves). The mode $\nu_{\uparrow\uparrow}$ is weakly voltage dependent.

Notably, the mode $\nu_\perp$ can be both integers and high fractional numbers, e.g. $\nu_\perp$= 1.9, 2,…5.1, 6.875, …9.1…, 23,…40, in the vicinity of $T_C$. The origin of the unusually high fractional $\nu_\perp$-numbers are significantly different numbers of edge modes, $\nu_1$ and $\nu_2$ [see Eqs.(4)], corresponding to the different carrier concentrations to the left ($n_1$) and to the right ($n_2$) of p-n junction boundary. The concentrations $n_1$ and $n_2$ are determined by the contributions of the gate voltage and polarization terms [see Eq.(3)], and so their difference, $n_1 - n_2 = 2\frac{P_S}{e}$, originates from the different direction of ferroelectric polarization, $-P_S$ and $+P_S$, in different domains. The difference $(n_1 - n_2)$ tends to zero with the disappearance of the spontaneous polarization in the paraelectric phase (i.e. above $T_C$).

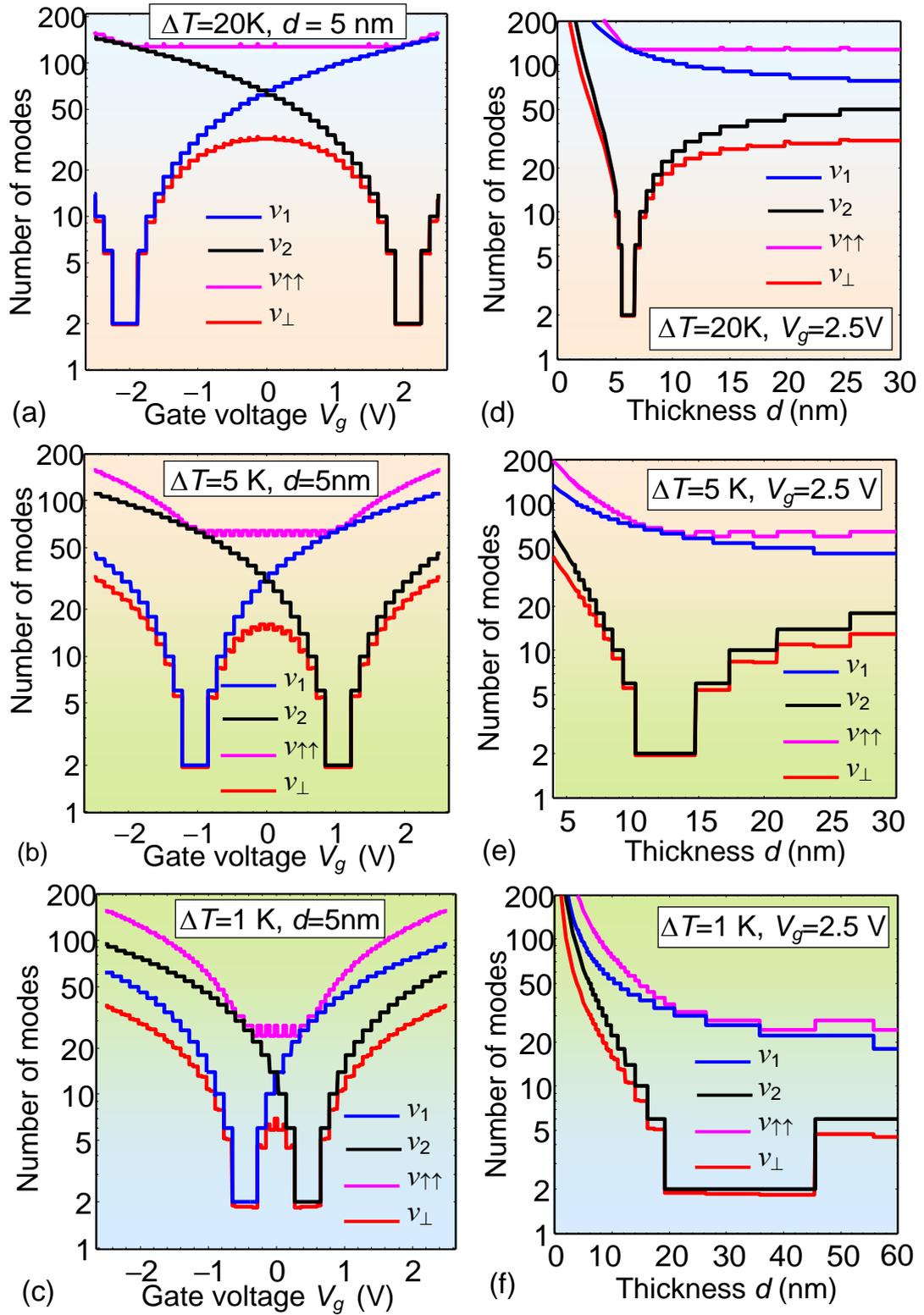

**Figure 2. (a-c)** The modes $|v_1|$ (blue curves), $|v_2|$ (black curves), $v_\perp$ (red curves) and $v_{\uparrow\uparrow}$ (magenta curves) in dependence on the gate voltage $V_g$ (in volts) calculated for $d = 5$ nm and the temperature difference $\Delta T = T_C - T$ equal to 20 K **(a),** 15 K **(b)** and 1 K **(c)**. **(d-f)** The modes $|v_1|$ (blue d curves),

$|v_2|$ (black curves), $v_\perp$ (red curves) and $v_{\uparrow\uparrow}$ (magenta curves) in dependence on oxide layer thickness $d$ (in nm) calculated for $V_g = 2.5$ V, the temperature difference $\Delta T = T_C - T$ equal to 20 K **(d)**, 5 K **(e)** and 1 K **(f)**. Other parameters $B$=10 Tesla, $\varepsilon = 7$, $T_C$=305 K and $P_0 = 0.1 \, C/m^2$.

The dependences of $|v_1|$, $|v_2|$, $v_\perp$ and $v_{\uparrow\uparrow}$ on oxide layer thickness $d$ are shown in **Fig.2(d)-(f)** for a fixed gate voltage $V_g = 2.5$ V. One pronounced stepped minimum is seen on the dependence $v_\perp(d)$ at the $d = \varepsilon\varepsilon_0 |V_g|/|P_S|$. The minimum position shifts from $d$=6 nm to 26 nm with the decrease of the difference $\Delta T = T_C - T$ from 20 K to 1 K [compare **Fig.2(d), (e)** and **(f)**]. The thickness dependences of either $|v_1|$ or $|v_2|$ are very close to the dependence $v_\perp(d)$ near the minimum (red curve almost coincide with either black curve or with blue one). The numbers of $|v_1|$ or $|v_2|$ are integers equal to 2, 4, 10, etc, but the number of perpendicular modes $v_\perp(d)$ appears fractional and its denominator differs from 2 or 3 with the increase of either $|v_1|$ or $|v_2|$.

Contour maps of the number of modes $|v_1|$, $|v_2|$, $v_\perp$ and $v_{\uparrow\uparrow}$ in coordinates "gate voltage $V_g$ – temperature $T$" are shown in **Fig. 3**. The maps of $|v_1|$, $|v_2|$ and $v_\perp$ contain the curved red contours on which the numbers are equal to "1" or "2" or close to them (e.g. 1.1, 1.9) and blue regions where the number is rather high [compare **Figs. 3(a), (b)** and **(c)**]. The high fractional number of $v_\perp$ happens when one of the edge modes (e.g. $|v_1|$) is minimal for the gate voltage close to the equality $|V_g| \cong |V_g^{cr}|$ in the condition (6), but another one (e.g. $|v_2|$) is far from it. The map of $v_\perp$ is symmetric with respect to the gate voltage sign [see **Fig. 3(c)**], and the maps of $|v_1|$ and $|v_2|$ are asymmetric and complementary to one another with respect to the voltage sign [compare **Figs. 3(a)** and **(b)**]. The number of $v_{\uparrow\uparrow}$ becomes relatively small only in the immediate vicinity of Curie temperature [see **Fig. 3(d)**].

Contour maps of the number of modes $|v_1|$, $|v_2|$, $v_\perp$ and $v_{\uparrow\uparrow}$ in coordinates "oxide layer thickness $d$ – temperature $T$" are shown in **Fig. 4** for a fixed gate voltage $V_g = 2.5$ V. The maps of $|v_1|$ and $v_{\uparrow\uparrow}$ have no features, but corresponding numbers change by an order of magnitude from tens to hundreds [see **Figs. 4(a)** and **(d)**]. The maps of the number of modes $|v_2|$ and $v_\perp$ look very similar and contain the curved red contours on which the numbers are small and blue regions where the numbers are rather high [see **Figs. 4(b)** and **4(c)**]. The number of $v_{\uparrow\uparrow}$ becomes smaller only in the immediate vicinity of Curie temperature at $d>20$ nm [see **Fig. 4(d)**].

From **Figs.3-4** the number of perpendicular modes $v_\perp$ varies from integers to high fractional numbers, e.g. $v_\perp$=1, 1.5, 1.6(6), 1.75, 1.9, 2,…5.1, 6.875, …9.1…, 23,…58, in the vicinity of transition from the ferroelectric to paraelectric phase.

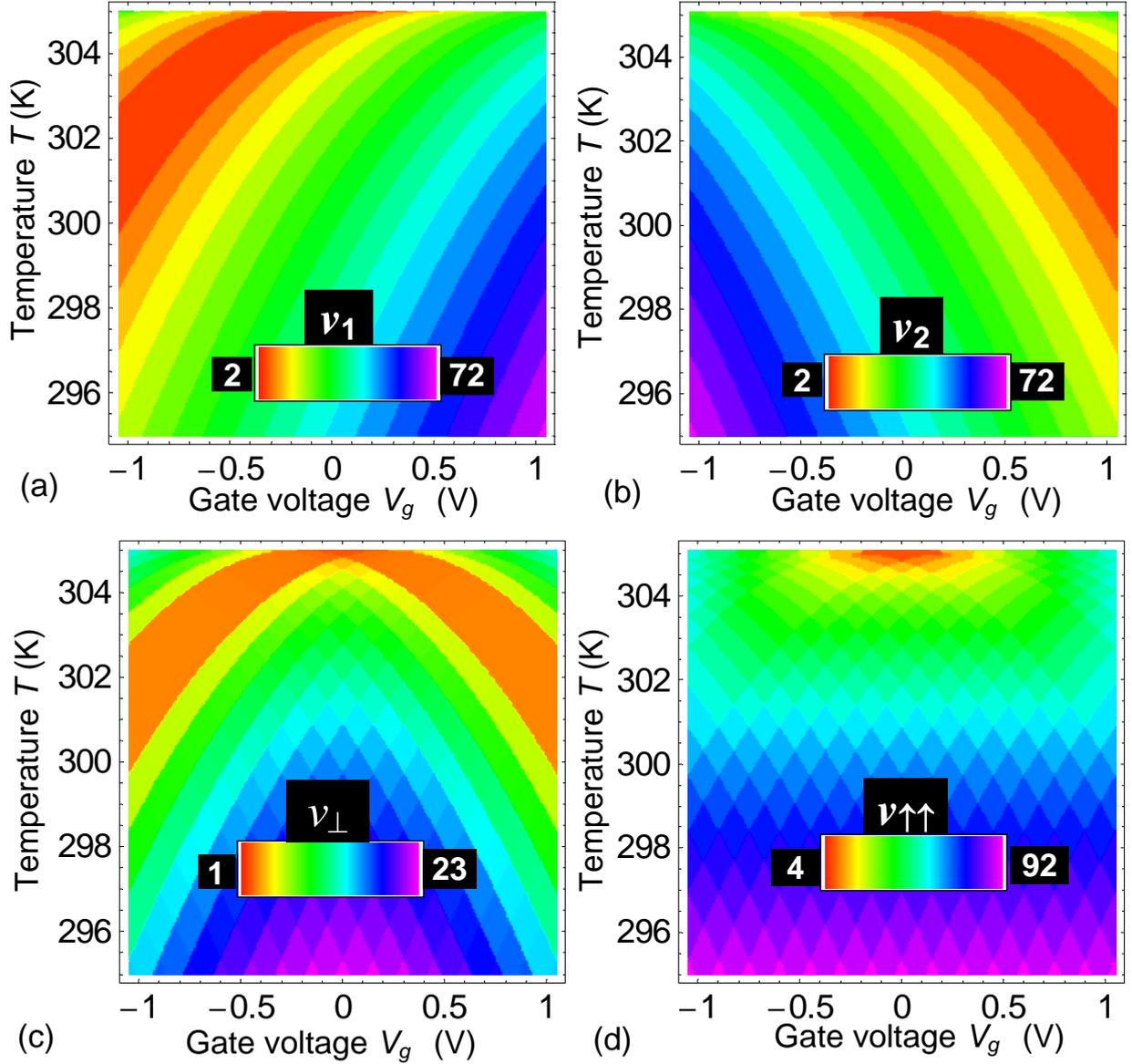

**Figure 3.** Contour maps of the modes $|v_1|$ **(a)**, $|v_2|$ **(b)**, $v_\perp$ **(c)** and $v_{\uparrow\uparrow}$ **(d)** in coordinates "gate voltage $V_g$ – temperature $T$" calculated for parameters $\varepsilon = 7$, $d = 6$ nm, $P_0 = 0.1$ C/m$^2$, $T_C$=305 K and $B$=10 Tesla.

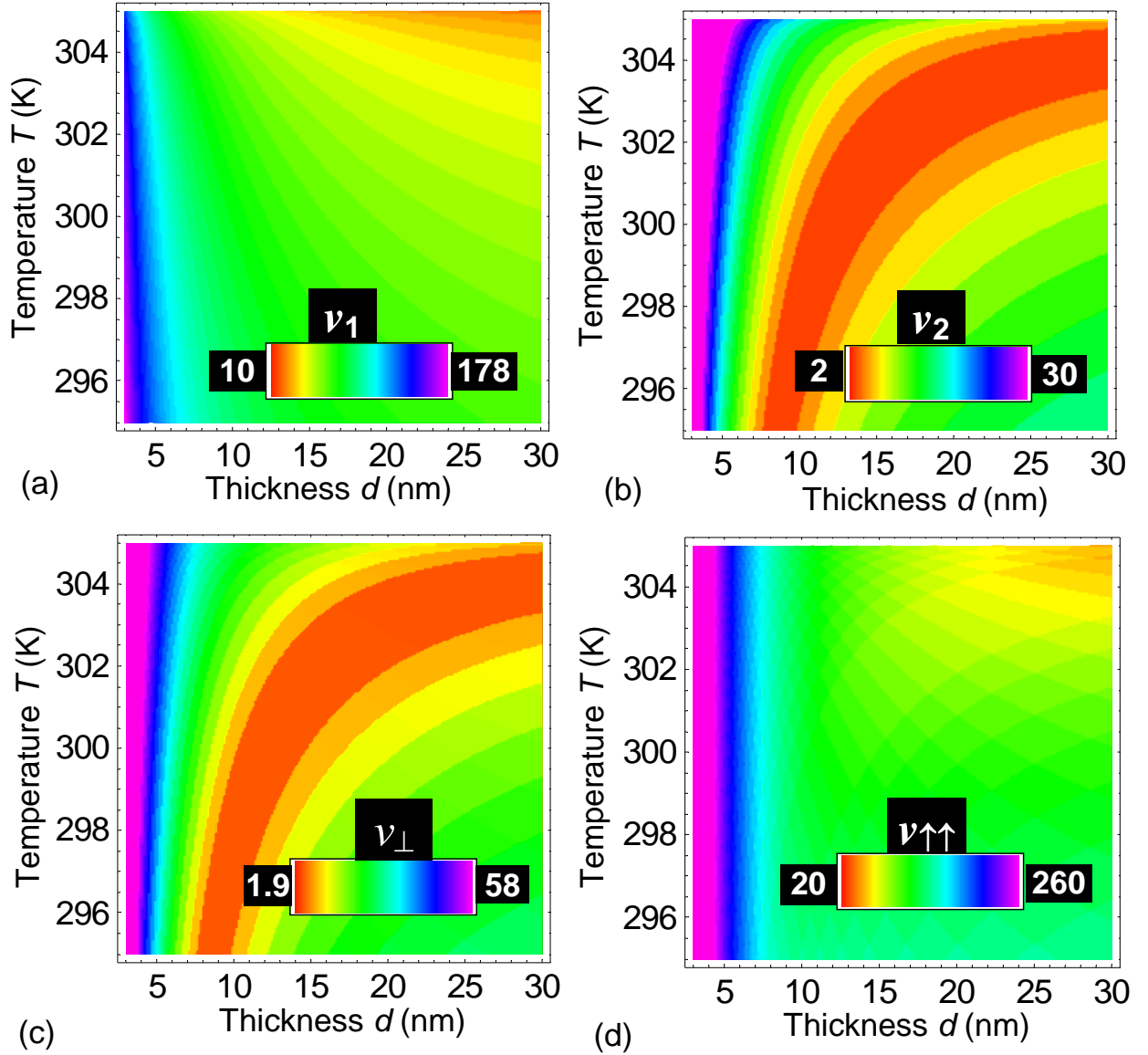

**Figure 4.** Contour maps of the modes $|v_1|$ **(a),** $|v_2|$ **(b),** $v_\perp$ **(c)** and $v_{\uparrow\uparrow}$ **(d)** in coordinates "oxide layer thickness $d$ – temperature $T$" calculated for parameters $\varepsilon = 7$, $|V_g| = 2.5$ V, $P_0 = 0.1$ C/m$^2$, $T_C$=305 K and $B$=10 Tesla. Color bars indicate the minimal and maximal numbers of the modes.

To resume obtained results illustrated by **Figs. 2-4,** we revealed that 180-degree FDWs in a ferroelectric substrate, which induce p-n junctions in a graphene channel, lead to nontrivial temperature and gate voltage dependences of the perpendicular and parallel modes of the unconventional integer quantum Hall effect. Unexpectedly the number of perpendicular modes $v_\perp$, corresponding to the p-n junction across the graphene conducting channel, varies from integers to various fractional numbers depending on the gate voltage, temperature and oxide layer thickness, e.g. $v_\perp$=1, 1.5, 1.6(6), 1.75, 1.9, 2,…5.1, 6.875, …9.1…, 23,…58, in the vicinity of transition from the ferroelectric to paraelectric phase. The great number of modes correspond the occupation of numerous LL in graphene placed of ferroelectric substrate with high spontaneous polarization, on the contrary low number of modes correspond the occupation of few LL.

These numbers of $v_\perp$ and their irregular sequence principally differ from the sequence of fractional numbers $v = 3/2, 5/3…$ reported earlier by Abanin at al [8]. The origin of the unusual $v_\perp$-numbers are the serial connection of different numbers of edge modes, $v_1$ and $v_2$ [see Eqs.(4)], corresponding to significantly different carrier concentrations to the left ($n_1$) and to the right ($n_2$) of p-n junction boundary. The concentrations $n_1$ and $n_2$ are determined by the contributions of the gate voltage and polarization terms [see Eq.(3)], and so their difference originates from the different direction of the spontaneous polarization, $-P_S$ and $+P_S$, in different ferroelectric domains. The strong difference disappears with the disappearance of the spontaneous polarization in the paraelectric phase (i.e. above the Curie temperature). The phase transition from the ferroelectric to paraelectric phase can take place either with the temperature increase (temperature-induced phase transition) or with the decrease of ferroelectric substrate thickness (thickness-induced phase transition).

These dependences of $v_\perp$ on the gate voltage and temperature reveal once more remarkable properties of mass-less Dirac fermions in graphene, and their nontrivial properties caused by ferroelectric substrate with domains can be promising for new types of modulators.

**Authors' contribution.** M.V.S. generated the research idea and performed analytical calculations. A.N.M. and A.I.K performed numerical calculations and generated figures. M.V.S. and A.N.M. wrote the manuscript.